# Positional Paper: Schema-First Application Telemetry

Yuri Shkuro, *Meta*    Benjamin Renard, *Meta*    Atul Singh, *Meta*

**ABSTRACT**

Application telemetry refers to measurements taken from software systems to assess their performance, availability, correctness, efficiency, and other aspects useful to operators, as well as to troubleshoot them when they behave abnormally. Many modern observability platforms support dimensional models of telemetry signals where the measurements are accompanied by additional dimensions used to identify either the resources described by the telemetry or the business-specific attributes of the activities (e.g., a customer identifier). However, most of these platforms lack any semantic understanding of the data, by not capturing any *metadata* about telemetry, from simple aspects such as units of measure or data types (treating all dimensions as strings) to more complex concepts such as purpose policies. This limits the ability of the platforms to provide a rich user experience, especially when dealing with different telemetry assets, for example, linking an anomaly in a time series with the corresponding subset of logs or traces, which requires semantic understanding of the dimensions in the respective data sets.

In this paper, we describe a schema-first approach to application telemetry that is being implemented at Meta. It allows the observability platforms to capture metadata about telemetry from the start and enables a wide range of functionalities, including compile-time input validation, multi-signal correlations and cross-filtering, and even privacy rules enforcement. We present a collection of design goals and demonstrate how schema-first approach provides better trade-offs than many of the existing solutions in the industry.

## 1. INTRODUCTION

Observability is a critical capability of today's *cloud native* software systems that power products such as Facebook, Gmail, WhatsApp, Twitter, Uber Rides, etc. Originally defined in control theory, observability provides operators with deeper insight into various aspects of the complex behavior of systems, including their performance, availability, correctness, and efficiency. When the systems behave abnormally, observability is used to troubleshoot the incidents and mitigate them to bring the behavior back to normal, with mean time to mitigation being one of the critical success measures.

To provide observability, the systems are instrumented to produce various telemetry signals. The most common types of application telemetry used with today's cloud native systems are metrics, logs, events, and traces [12], [21]. A common characteristic of different telemetry types is that they usually combine one or more measurements with a set of identifying dimensions. For example, a metric is a numeric observation typically associated with a name, such as "request_count", and some dimensions, such as "host" or "endpoint". Similarly, in a semi-structured log message, the measurement part is played by the message text, accompanied by searchable dimensions such as log level, thread name, etc.

Modern telemetry platforms, in addition to ingesting vast amounts of telemetry data, usually perform extensive indexing of the dimensions to allow rich querying and aggregations over the raw measurements [17], [10], [2]. Most of them treat dimensions as free-form collections of key-value pairs. Platforms like OpenTelemetry [15] or Jaeger [20] allow associating basic types with dimension values, while systems like Prometheus [6] allow associating descriptions with the metrics while treating all dimensions as strings. Little, if any, additional metadata is captured or understood by these systems. This puts a burden on the user to understand how to interpret the dimensions and how to leverage them when querying data.

The complex nature of cloud native systems often requires investigations that involve more than a single source of telemetry. A spike in error rate in a single zone might warrant a look at the logs or traces from the same zone for better diagnosis of the issue. This is where many modern telemetry platforms fall short, as they lack semantic understanding of the data. Two telemetry signals might share a dimension "region", but in one case referring to the region where the software runs and in the other case to the region where the user is located. Joining telemetry by this dimension as if it is the same thing is probably meaningless. Metadata can be the missing link in solving these problems.

In this paper we define *metadata* as additional information that provides semantic meaning to telemetry data and helps in identifying the nature and features of the data. Examples of observability metadata include data types, units, descriptions, ownership, purpose policies, semantic identifiers, etc.

There are different ways to associate metadata with telemetry, such as using naming conventions to imply semantic meaning or defining metadata *a-posteriori*, after the telemetry data has been produced and stored. In this paper we propose a schema-first approach to capturing metadata for application telemetry that we believe provides better trade-offs compared to other solutions. Schema-first development is a well-known technique, especially in the areas of data management and API design. However, it is rarely used with application telemetry because it can create significant friction to developer experience compared to the simplicity of conventional telemetry APIs, like logging a message or incrementing a counter in a single line of code.

This paper makes the following contributions:

- Provide analysis of the existing approaches to telemetry metadata that are prevalent in the industry.
- Demonstrate how schema-first approach can be applied to telemetry by addressing the usability and change management issues.
- Propose a comparison methodology for evaluating metadata approaches.

The rest of the paper is organized as follows. Section 2 describes our motivations for associating metadata with application telemetry. In Section 3 we present our design goals and evaluation criteria. In Section 4 we review the existing approaches from the industry for associating metadata with telemetry. In Section 5 we present our schema-first solution, discuss its implementation details and features, and evaluate it against our design goals. And in Section 6 we discuss the conclusions, applicability of the approach to other situations, and future work.

## 2. MOTIVATION FOR METADATA

Metadata in application telemetry is a means to an end: the users of observability platforms do not consume metadata directly. However, it is an essential building block to solving higher-level problems.

**Discoverability of data**: defined as the ability to find the relevant telemetry assets at the right time. Discovery has a dependency on metadata to answer search queries based on metadata (e.g., what are all the telemetry artifacts owned by my team?), and to provide semantic information, such as human-readable descriptions, so that users can validate that the results of the search query match their needs.

**Exploration of data**: once a user has discovered the right asset, they need to explore it to extract the information they are looking for. Exploration has a dependency on metadata (a) to determine which operations are allowed on a particular asset, e.g., which dimensions are available for filtering or grouping, and (b) to allow exploration of data across assets, e.g., by joining or cross-filtering on shared dimensions.

**Investigative assistance**: the ability to automatically extract insights that can help users speed up an investigation. As a form of automated data exploration, it depends on metadata to have a semantic understanding of the data, for example to know which dimension is a region or whether that region is a source or destination region in a message transmission.

**Privacy**: while application telemetry is generally not meant to contain sensitive user data that may be subject to privacy policies, it may be possible for the applications to leak sensitive data into telemetry by accident, or sometimes to include it intentionally with the expectation of certain access controls. Metadata helps the systems understand which parts of the telemetry assets may contain sensitive data, what ownership and access controls exist for this data, what policies may govern the retention of this data, and how lineage tooling may be used to automatically identify sensitive data both entering the telemetry data streams or being transformed into other aggregate data sets.

## 3. EVALUATION CRITERIA

When analyzing the costs and benefits of different approaches to telemetry metadata, we found the following design goals and evaluation criteria to be important to engineers at Meta.

**Design considerations.**

- C0: does the approach encourage engineers to think through the implication of adding new data to telemetry, such as whether it is privacy-sensitive, or whether the semantic type of the data already exists somewhere and should be reused for possible cross-asset correlations?

**Authoring experience**. Most of the existing telemetry APIs are designed to make it as simple as possible for developers to log telemetry data points. Most logging frameworks support the simplicity of a `printf` statement; many metrics libraries allow emitting a new counter with a single line of code. This simplicity and low friction are very important because if it is cumbersome for engineers to add instrumentation to the code, we end up with code that emits no telemetry and provides no observability. We consider the following criteria in this category:

- C1: does the approach require more lines of code to emit telemetry?
- C2: does the approach make it more difficult to deploy a change, e.g., by requiring the developer to run an equivalent of ALTER TABLE command before the code hits production?
- C3: how does the approach affect distributed authoring workflow when multiple teams own different parts of a data set?
- C4: can the solution enforce schema consistency across different log sites? For example, preventing Java and Go programs from emitting the same type of time series with incompatible shapes.

**Change management**. While many telemetry APIs are designed to support free-form dimensions on telemetry, this is not a full picture of the life cycle of telemetry data. Once an application emits a time series, it can have many consumers for that data, from automated detectors analyzing the time series for anomalies to visual dashboards configured with a certain understanding of the available dimensions. Changing the shape of the emitted telemetry, such as adding, removing, or renaming a dimension, can easily break those consumers, so we pay attention to change management practices that a given approach to metadata affords.

- C5: does the approach allow evolution of telemetry shapes and schemas over time, such as the need to sometimes rename fields?
- C6: can the approach automatically identify breaking changes to telemetry, e.g., in the form of continuous integration checks?
- C7: can the approach provide compile-time safety against incompatible changes, such as supplying semantically incompatible value to a dimension, e.g., a fully qualified host name instead of a short host name?
- C8: does the approach support automated code changes (often called "codemods" at Meta), like renaming a column in both producers and consumers?

**Querying**.

- C9: does the collected metadata allow automated introspection of telemetry assets and presenting to users only the choices that are applicable?
- C10: does the metadata allow understanding of semantically identical dimensions across data sets, to enable cross-asset filtering and consistent querying?

## 4. EXISTING APPROACHES

The need for telemetry metadata is well understood, evidenced by solutions going back more than a decade. In this section we evaluate some of the popular existing solutions against our design goals. We mostly limit the discussion to open-source observability products.

### 4.1. STATSD

The Statsd protocol [13] became very popular for system metrics due to its simplicity and plain-text exposition format. It provided little in the way of capturing metadata of the metrics, restricting the observation to a single string name and a value associated with a type like "counter" or "gauge". The data model did not provide any dimensional support, but in the existing backends, such as Graphite [4], the metric name was understood to be structured as a collection of dot-separated segments, which were used by operators to encode interesting dimensions into the metric name. For example, if we wanted to count the number of requests received by a service and further partition this time series by dimensions like service name, protocol, and status code, we could encode it as `{service}.reqs.{protocol}.{status}`. The backends explicitly supported query functions that understood such notation, e.g., to aggregate all metrics for all services but partition them by the status code we could use a function `groupBy(4)`, which refers to the fourth segment in the metric name.

It is obvious that Statsd protocol meets none of our metadata requirements. In particular, the lack of true dimensional model made writing queries on Statsd metrics very unintuitive because users needed to refer to segments by their index, which also made queries very easy to break by changes in the way metrics were produced.

### 4.2. DIMENSIONAL MODELS

Seeing the wide adoption of Statsd protocol, the industry practitioners wanted to improve the dimensional aspects of the model. Google's Monarch [2], Uber's M3 [19], and Prometheus [6] are examples of metrics backends that explicitly support dimensional data models where a metric name can be associated with a group of string key-value pairs (often referred to as labels or tags), such that users could write queries explicitly referring to dimensions by name, e.g., `http_requests{job="foo", group="canary"}` in Prometheus query language.

Similar free-form dimensional models are supported by other telemetry platforms, including the open-source distributed tracing systems Jaeger [20] and Zipkin [3], as well as instrumentation-oriented projects like OpenTelemetry [15]. In OpenTelemetry, every telemetry asset (metrics, traces, and structured logs) can be associated with *attributes*, which are still mostly free-form key-value pairs, except that the values have types, either primitive (string, number, Boolean) or complex types built with nested arrays and maps.

Named dimensions were a vast improvement over Statsd model, resulting in much more intuitive query expressions. The queries are also more resilient to upstream changes in the telemetry signals, because the order of dimensions is irrelevant. Yet the free-form format of the dimensions is still far from meeting our requirements for metadata. Essentially, all these systems adopt a *code-first approach*, where the code producing the telemetry is the final authority on the schema, shape, and semantic meaning of the data, yet none of this metadata is captured or made available to the consumers. When users query the data, they need to find out, through other means, which dimensions are present in the telemetry assets, what those dimensions are called, and what values are allowed. Usually, this leads to large inconsistencies between telemetry data produced by different components. Change management is also very complicated; it is very easy to break the consumers by changing the producing code, and the framework provides no mechanism for a feedback loop.

### 4.3. SEMANTIC CONVENTIONS

To impose more structure on the telemetry data, different projects define *semantic convention* that prescribe how certain common dimensions should be named in the telemetry and which values can be assign to them. One such example is the Elastic Common Schema (ECS) [8], an open-source specification that defines a common set of fields to be used when storing event data, such as logs and metrics, in Elasticsearch. ECS specifies field names and datatypes for each field and provides descriptions and sample usage. For example, instead of dealing with potentially many ways of representing a source IP address in different data sets, the consumers of ECS-compliant data can rely on this dimension always be called `source.ip`.

Similar mechanism exists in the OpenTelemetry project called *semantic conventions* [16]. As of v1.9, the conventions are defined for metrics, traces, and resources (*resource* is a software component whose behavior is described by the telemetry, e.g., a host, a process, or a Kubernetes cluster). Similar to ECS, the OpenTelemetry semantic conventions specify dimension names, their descriptions and semantic meaning, units of measure, and value types. This metadata is formally encoded in the YAML files that are a part of the OpenTelemetry specification.

The OpenTelemetry attribute names also use dot-separated notation, but it is treated differently than in the Statsd protocol. The dot-segments are not used to encode dimension values, only to represent dimension namespaces and to group attributes by some common characteristic. For example, all keys of the form `net.*` refer to network-related attributes in general, while all keys of the form `net.peer.*` refer to network attributes of a remote peer communicating with the component producing telemetry. Concrete examples of the keys are `net.peer.ip` and `net.peer.port`, which respectively refer to the IP address and port of the remote peer.

Semantic conventions for telemetry address some of our design goals for metadata. Producers have well-defined expectations for the shape of the telemetry dimensions and for using compatible values across different assets. Consumers can rely on the conventions to know which dimensions they can use when querying the data. The authoring experience of generating compatible telemetry is not particularly burdensome, especially when language implementations expose static constants for different attributes defined in the semantic conventions.

The main downside of the semantic conventions approach is that they are, after all, only conventions. There is no built-in mechanism to guarantee that the conventions are used correctly and consistently. There is no strict type and value checking, for example, to ensure that a field that is meant to contain milliseconds is not assigned a value in seconds. There is no systematic way of warning developers at coding time if they are going to break consumers by changing the shape of the data.

Another downside is the difficulty of retrofitting existing data sets to match the new conventions. It requires either full migration of the producers and consumers of the data set to a new format, which is usually cost prohibitive, or introducing a data transformation process into ingestion pipelines, which is a more common approach, but it incurs performance overhead and a long-term maintenance burden.

### 4.4. OPENTELEMETRY SCHEMAS

The OpenTelemetry authors understood that change management is an important aspect of the framework. As of v1.8, the OpenTelemetry Specification introduced the notion of *telemetry schemas* [14]. The schemas, despite the name, do not actually provide a formal definition of the telemetry attributes, as those are already defined by the semantic conventions themselves. Instead, schemas support versioning of the semantic conventions by describing the changes between versions, specifically renaming of the attributes, with more transformations possibly supported in the future. The emitted telemetry is expected to contain a URI, such as `https://opentelemetry.io/schemas/1.9.0`, referring to the schema and its version used by the producer. The telemetry backend can use that to automatically upgrade or downgrade the telemetry representation for the producers or consumers that are using an older version of the specification.

Similar methods are used by other specifications. ECS defines a field `ecs.version` [9] that encodes the version of the specification employed by the ingestion pipeline. The CloudEvents specification [7] uses a `dataschema` field to encode the URI of the schema used by the events. Both ECS and CloudEvents only allow identifying which version of the schema is used by a specific instance of telemetry data, but do not describe the mechanism for transformation between versions that is possible with the OpenTelemetry schemas.

While knowing the version of the schema used by a telemetry asset is important, it does not fundamentally change the limitation of the semantic conventions approach that we discussed previously.

### 4.5. EXTERNALLY AUTHORED METADATA

So far, we discussed the approaches that attempt to introduce some *a-priori* knowledge of the metadata, before the telemetry is produced. The alternative to that is an *a-posteriori* enrichment where the metadata is defined after the telemetry is produced and captured in the storage backends.

This approach requires a designation of some system as an authoritative source of metadata about various data sets, a *metadata store*. The consumers of telemetry can consult the metadata store to know how to interpret the data. At Meta, the concept of metadata store has been used for several years, especially for the vast amounts of data stored in the data warehouse.

The metadata authored externally in the metadata store addresses many of our querying requirements but falls short in other areas. One of its drawbacks is the lack of consistency between the telemetry authoring and the metadata, which means the actual schema of the produced telemetry can easily get out of sync with the metadata, since the latter needs to be proactively updated after the fact (although some automation can be helpful for this). Another challenge is poorly defined identity of the telemetry assets. A metadata store requires a unique identifier with which it can associate the metadata, but not all telemetry types are able to provide that. For instance, a stream of structured events written to a Kafka [18] stream or a Scribe [11] category may be uniquely identified by the name of the destination stream/category, but a certain shape of data logged by a specific microservice into a distributed tracing platform like Jaeger [20] or Canopy [10] has no well-defined identity to which a metadata can be attached a-posteriori.

### 4.6. AUTOMATIC DATA ENRICHMENT

Certain classes of telemetry can be automatically enriched by the publishing libraries or the collection pipelines with standardized dimensions describing the resources, based on the inherent knowledge of the underlying infrastructure. For example, an application may emit a time series `requests_total_count` without providing any additional dimensions, and the collection pipeline can automatically add attributes like service ID, host name, pod name, zone, etc. This approach is widely used by commercial Observability vendors whose collection pipelines and agents support integrations with dozens of popular infrastructure components. Because the ownership of all these enriched dimensions is centralized, the Observability platforms can provide consistent view of telemetry metadata to consumers. Automatic data enrichment is complimentary to the schema-first approach we propose, and can itself be implemented as schema-first, but in isolation it does not fully meet our requirements because it does not help with any custom, non-infrastructure related dimensions that applications often want to use with the telemetry.

### 5. SCHEMA-FIRST TELEMETRY

After evaluating many options, we concluded that a *schema-first* approach to application telemetry will be the most beneficial for engineers at Meta, and our team is currently working on building the necessary tooling to support it. In contrast to the code-first approach that is most prevalent in the observability industry, schema-first means that the design of the new telemetry assets starts with the schema; the schema is formally specified using some interface definition language (IDL, in our case, Thrift); and this schema becomes the single source of truth about the metadata of the asset. In this section we describe some of the implementation details of this approach and evaluate it against our design criteria.

Schema-first approach to data management is not a new concept for engineers at Meta, especially for business analytics data. A typical path for this type of data starts with structured events being logged into Scribe [11], from where they are ingested either into Scuba [1] for real-time analytics or into the data warehouse for batch processing. Historically, the events were written to Scribe using a code-first, free-form API, which only captured the types of the fields but no other metadata, and the real metadata was then curated in the metadata store. Because this approach, as we discussed earlier, is prone to emerging inconsistencies between the metadata and the data itself, the company developed a schema-first logging framework to which many important data sets have been migrated, resulting in much better stability and reliability, improved controls and privacy, automated lineage, and better efficiency. We are extending that approach to all types of application telemetry by reusing many of the building blocks of the schema-first logging framework.

### 5.1. IMPLEMENTATION

Perhaps the biggest trade-off in the schema-first approach is the introduction of extra steps in the authoring process. Our goal was to minimize this impact, but we were not able to eliminate it completely. In the end, we consider this a worthy trade-off given the other benefits that we get from strong metadata support. It is also helpful that many engineers at Meta are already familiar with the proposed workflow.

Let us first consider the base case of adding an extra piece of data to an existing telemetry asset, e.g., a shard ID dimension to the request counter in an RPC server. In the code-first approach this requires adding a single line (Listing 1). In the schema-first approach, this counter would already have a schema defined using an IDL (Listing 2) and the application code to increment the counter gets one extra line (Listing 3). Compared to the code-first approach, the schema-first solution requires at least one extra line of IDL code to define a new field, or even more lines if we want to provide additional metadata, in this example by defining a custom type `ShardID` that later can allow stricter validation of the values.

It is worth noting how the schema-first approach provides potential efficiency improvements: in Listing 1 the dimensions are passed as a map, so the keys will have to be included in the wire format, while in Listing 3 we are using a strongly typed data structure that can be efficiently serialized into a binary Thrift payload.

```
counter.Increment(
  service_id  = 'foo',
  endpoint    = 'bar',
  status_code = response.code,
  shard_id    = 'baz',   // added line
)
```
Listing 1. Adding new dimension with traditional API.

```
typedef string ServiceID
typedef i32    StatusCode
typedef string ShardID    // added line

struct RequestCounter {
  1: ServiceID  service_id
  2: string     endpoint
  3: StatusCode status_code
  4: ShardID    shard_id // added line
}
```
Listing 2. Adding new dimension to the schema.

```
counter.Increment(RequestCounter(
  service_id  = 'foo',
  endpoint    = 'bar',
  status_code = response.code,
  shard_id    = 'baz',   // added line
))
```
Listing 3. Emitting new dimension via a struct.

Besides the extra lines of code, we can also see that the schema-first solution requires some developer tooling support. First, the RequestCounter type is auto-generated from the Thrift definition. At Meta, this is a standard and fully automated process, because the Buck build tool [5] knows how to handle Thrift IDL files. Second, if the telemetry is indeed serialized on the wire using Thrift binary format, then the consumers of this data must have access to the schema before they can parse the payload. This part has been already solved by Meta's schema-first logging framework we mentioned earlier, which implements a schema actualization process executed automatically when the code change passes the continuous integration (CI) tests and is merged into the main branch. Actualization involves validation of the schema changes against backwards-incompatible changes (such as changing a field type). If the validation fails, the code change fails the CI and will not be merged. If the validation is successful, then the schema change is pushed as a new version to the metadata store, where it can be distributed to real-time consumers such that when the new shape of telemetry is produced in production the consumers already have the new schema and are able to parse the data.

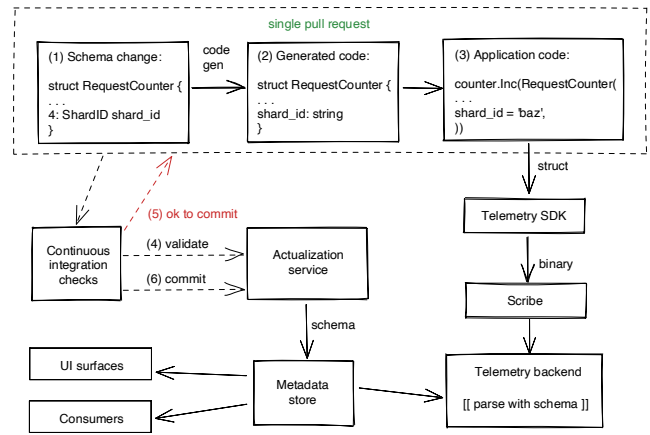

Figure 1. Schema-first authoring process.

Figure 1 shows the overall process of authoring and deploying a change to telemetry. Both the schema change (1) and the application code change (3) can be done in a single merge request, an important attribute of the solution for a better developer experience. CI checks integrate with actualization service to validate and update the schema in the metadata store. The telemetry is serialized by the SDK into a binary payload on the wire (or into Scribe), which is deserialized by the telemetry backend having runtime access to the updated schema. Consumers and Observability tools can consult the metadata store when discovering or accessing the data.

For authoring new telemetry assets, the process requires more work on the schema part, that may involve such steps as: (a) creating a new IDL file with new data type, (b) adding a dependency reference to the application build file to let the build system know that Thrift code generation step is required before building the application, (c) importing the generated package into the application code to make the new data type accessible in the code. We envision that these steps can be easily automated with a wizard-like command line tool or a build target.

There are several reasons why we chose Thrift as the language for defining the schemas. Thrift is the de-facto standard at Meta for defining interfaces between services, so most engineers are already familiar with it. It has strong tooling and IDE support for authoring, code-generation, and automatic cross-repository syncing. The code generation and serialization are well supported across most languages used in the company. The IDL itself is language-neutral and very expressive, with support for type aliases and annotations, that are critical for the rich metadata needs discussed in the next section. Thrift also supports namespaces and composition, which is important for the reuse of data types and semantic annotations, as well as for collaborative authoring of large data sets that are shared across multiple teams.

Special handling is needed when trying to apply schemas to distributed tracing data [21]. The other types of telemetry usually have a well-defined identity and owners and are isolated from each other, e.g., a stream of structured logs from a service can have its own schema that is independent of all other telemetry assets; similarly, a set of metrics produced by an RPC framework can be represented by a schema owned by the library authors. In contrast, a single trace contains information collected across many different services. Some of the data in the trace could follow a common schema (such as the infrastructure dimensions that describe a service), while other data could be unique to each participating service.

Traces are usually modeled as a collection of spans, where a span represents a certain operation performed by the software. The granularity of the operation is arbitrary; it could represent the whole RPC request handled by a service, or it could wrap a single function call within the code. A span provides a single generic envelope for capturing debugging and performance data, and it is up to the application code to decide which data to store there. In order to keep this open-ended nature of tracing data yet still allow strong schematization of the data, our approach is to allow different teams to define the schema for their portion of the data and their use case, and extend the span model to store these schematized binary payloads keyed by the fully qualified name of the schema data type, e.g., in a `string→[]byte` dictionary. For example, the Canopy tracing platform provides an out-of-box instrumentation for our standard RPC framework. The tracing team can define a common schema for the way we want to capture data about RPCs; the instrumentation will populate this data type and store it in the trace under `canopy.core.rpc` key. Meanwhile, an individual service team X that wants to capture some data specific to their service can define another schema and store the data with `serviceX.dataY` key. This way we preserve the existing flexibility of traces being able to store any kind of data, while enjoying all the benefits of the schema-first approach. On the reading side, the consumers can access the data via strongly typed data structures auto-generated from the schema. In cases where more dynamic access is needed, e.g., in a generic layer that transfers the data from a queue into the data warehouse, the data can be parsed using the runtime schema information available from the metadata store.

```
struct HostResource {
  @DisplayName{"Host ID"}
  @Description{"Unique host ID.
    For Cloud, this must be the
    instance_id assigned by the
    cloud provider."}
  1: string id

  @DisplayName{"Hostname (short)"}
  @Description{"Name of the host as
    returned by the 'hostname'command."}
  2: string name

  @DisplayName{"Architecture"}
  @Description{"The CPU architecture
    the host system is running on."}
  3: string arch
}
```

Listing 4. Schema-first model for OpenTelemetry semantic conventions for host resources.

```
// Example: devvm123
@DisplayName{name="HostName"}
@SemanticType{InfraEnum.DataCenter_Host}
typedef string HostName

// Example: devvm123.zone1.facebook.com
@DisplayName{name="HostName (with FQDN)"}
@SemanticType{InfraEnum.DataCenter_Host}
typedef string HostNameWithFQDN
```

Listing 5. Defining two different representations of the same semantic type.

```
enum OneWayMsgExchangeActorEnum {
  SOURCE = 1, TARGET = 2,
}
@SemanticQualifier
struct OneWayMsgExchangeActor {
  1: OneWayMsgExchangeActorEnum value
}
struct RPC {
  @OneWayMsgExchangeActor{SOURCE}
  @DisplayName{"Source service"}
  1: ServiceID source_service

  @OneWayMsgExchangeActor{TARGET}
  @DisplayName{"Target service"}
  2: ServiceID target_service
}
```

Listing 6. Qualifying rich type fields with additional semantic meaning.

## 5.2. RICH METADATA

The basic workflow we described so far makes the solution roughly equivalent to the code-first approach with a-posteriori metadata. However, once we have a data type described in the schema file, we have a strong foundation to start adding more metadata that can meet our design goals, using annotations and rich types.

The version of Thrift used at Meta supports adding metadata to the basic schema definitions using *annotations*, which are similar to annotations in the Java language. Annotations themselves are first declared as data types with fields, and then used to annotate the other data. For example, we can annotate fields in a struct with additional metadata, such as description or a display name. Listing 4 shows how the OpenTelemetry semantic conventions for a host resource could be represented in the schema.

The annotations mechanism allows us to describe very rich metadata about the fields, such as the types of query operators that should be allowed for a field, or the validation rules for the input data (e.g., a regular expression). The annotations can be used to describe not only the dimensions, but the measurements as well, e.g., to emphasize that a numeric field represents duration in milliseconds rather than seconds.

The *rich types* allow us to take this even further and assign semantic meaning to the fields. Notice how the "name" field in Listing 4 is described as the short form of the host name, as opposed to the fully qualified name that includes the domain and sub-domains. This information is only encoded in the metadata intended for human readers. Instead, we can define `HostName` as an alias to a string type that can capture this additional semantics (Listing 5) and use it as the data type for the name field in the `HostResource` type.

Listing 5 illustrates several additional benefits of this approach. The two rich types share the same semantic type, `DataCenter_Host`, which allows the consumers to recognize that even though two telemetry assets may have different representation of the host name field, they nonetheless refer to the same kind of data, so that, for example, we could apply host name filter to both assets simultaneously when querying or plotting the data. On the other hand, the short-form host name may not have a one-to-one mapping to the fully qualified host names, which is a fact that can also be encoded in the annotation on the rich types. In general, we can use annotations to describe detailed relationships between the types, such as how to convert values from one representation to another.

The other significant benefit is that these definitions of rich types can be easily reused across different telemetry schemas. In our example the description and display name annotations are defined on the rich type itself, so that they do not need to be repeated across field definitions using this type, which provides more consistent experience to the consumers of the data.

The semantic annotations allow encoding custom meaning of the fields in different data sets. For example, let us assume that we defined rich semantic types for such dimensions as region, host, service. Now consider a stream of events or metrics produced by a service mesh about the RPC activity between microservices. It is not enough to have rich types for region, host, and service, because an RPC involves two of each, a source and a target. If we simply encode this aspect in the field names, such as `source.service` and `target.service`, we are back to the semantic conventions situations that relies on the naming of fields. However, with annotations we can attach this metadata explicitly in machine-accessible way. In Listing 6 we add a new annotation type `OneWayMsgExchangeActor` that allows us to distinguish between the source and target service IDs at the metadata level, not just the field name level. The new annotation is itself tagged as `SemanticQualifier` to allow the platform to recognize it as special type of metadata.

Finally, the annotations can be used to capture privacy-related metadata on the fields, such as owners and required access controls, retention policies, purpose and allowed usage policies, etc.

## 5.3. EVALUATION

We can now evaluate the proposed schema-first approach against our design goals.

**Design considerations**

- C0: the schema-first approach encourages engineers to think through the implication of adding new data to telemetry, such as whether it is privacy-sensitive, or whether the semantic type of the data already exists somewhere and should be reused for possible cross-asset correlations.

**Authoring experience**

- C1: the approach requires slightly more code (usually one extra line in the IDL) to emit telemetry compared to the code-first approach.
- C2: deploying a change is not more complicated provided that the necessary automation exists.
- C3: the Thrift IDL approach can scale when multiple teams own different parts of the same data set, by using composition of type definitions. It may require minor enhancements to the Thrift compiler to support flattening of nested structs into a single struct (similar to struct embedding in the Go language).

- C4: emitting telemetry is done by interacting with the datatypes auto-generated from the schema definitions, which provides consistent authoring experience and shape of data, even when the emitting code is in different languages or located in different source repositories. In case of different repositories, extending the shape of telemetry may require more than one merge request, one to author the actual schema change in the repository where the primary schema source is defined, and the other to sync the changes to the schema to another repo (a workflow that is fully automated at Meta).

**Change management**

- C5: the approach is designed to allow evolution of telemetry shapes and schemas over time. The schema actualization service ensures that the schema changes are backwards compatible.
- C6: in addition to guarding against breaking changes purely at the schema level, the approach can also warn about breaking the consumers that depend on specific shape of telemetry. Specifically, when consumers use datatypes auto-generated from Thrift IDL, we can use static code analysis or lineage frameworks to identify affected consumers.
- C7: since the application code populates the telemetry by invoking setter functions on the datatypes auto-generated from the Thrift IDL, these functions provide a leverage point for enforcing data validation rules, either at runtime, or at compile time by using strong types (if at all possible, in a given language).
- C8: the approach is theoretically compatible with automated code changes, but we have not yet invested the time to integrate it with the existing *codemod* frameworks.

**Querying**

- C9: all changes to the schema end up in the metadata store, from where they can be retrieved by the tools and consumers to allow automated introspection.
- C10: rich types and semantic annotations in the schema provide the consumers with semantic understanding of the telemetry data and its dimensions, and support building rich functionality, from introspection and cross-asset filtering to machine-learning based analysis.

## 6. DISCUSSION

Access to metadata about application telemetry has many benefits across the whole data pipeline, from producers to consumers, with use cases ranging from data validation and change management to cross-dataset correlations and privacy enforcements. There are many projects in the industry that are trying to associate metadata with telemetry, usually by means of semantic conventions and schemas. This paper introduces a schema-first approach to capturing telemetry metadata that has not found its place in the observability industry so far. We described its implementation details and demonstrated that it provides better trade-offs than many of the existing solutions.

The schema-first approach to application telemetry described in this paper is aspirational and a work-in-progress at Meta. While it is based on the existing solutions in the company for business data, it has not been applied at scale to system telemetry. Our analysis indicated that it can be implemented at Meta, but we do recognize that it has certain trade-offs, namely in making the authoring developer experience more complicated, and these trade-offs may not be acceptable in different organizations and engineering cultures.

We contrasted this approach with many other solutions that exist in the industry. Since some of the paper authors are closely involved in the open-source projects like OpenTelemetry, a reasonable question is whether the schema-first approach should be attempted there. We hesitate to make such a recommendation at this time, because we still want to prove the approach at scale at Meta, and because the approach requires a significant level of consolidation and enhancements in the tooling specific to a certain developer experience ecosystem. For example, the centerpiece of the approach is authoring of schemas in Thrift IDL, which is only one of the competing interface languages in the industry. Open-source projects like OpenTelemetry aim to be applicable across many different technologies, so it is difficult for them to take a hard dependency on any specific IDL.

There are areas of the schema-first solution that require additional design decisions and future work. We do not have a fully developed framework for versioning and A/B testing. For example, if the schema changes introduced in a merge request are automatically pushed to the metadata store, how can a merge request be tested in a staging environment or canary deployments before rolling out to production? The schema change may be backwards compatible, but if it needs to be reverted, it could leave the emitted telemetry data in an inconsistent state. This problem is not unique to our solution, since any data store with schematized data must deal with this.

Another area that will require attention is data governance mechanisms. The success of our approach is predicated on good reuse of semantic annotations in the schema, to allow the consumers to see similarities between data sets. We are too early in the process to decide how such reuse will be achieved. We are starting with developing a curated set of data types for common infrastructure dimensions that are shared by many telemetry data sets (the OpenTelemetry semantic conventions provide a good starting point for this).